\documentclass[numberedappendix,aps]{emulateapj}
\usepackage{amsmath}
\usepackage{newtxtext}
\usepackage[varg]{newtxmath}
\usepackage{natbib}
\usepackage{graphicx}
\usepackage[svgnames]{xcolor}
\usepackage{color}
\usepackage{multirow}
\usepackage{array}
\usepackage[hyperindex,breaklinks]{hyperref}

\hypersetup{
  breaklinks=true,   
  colorlinks=true,   
  pdfusetitle=true,  
  citecolor=blue,
  urlcolor=blue,
  linkcolor=blue,
}

\newcommand{\Msun}{M_\odot}

\newcommand{\nvis}{n_{\rm{vis}}}
\newcommand{\Hvis}{H_{\rm{vis}}}
\newcommand{\avis}{\alpha_{\rm{vis}}}
\newcommand{\tcol}{t_{\rm{BH}}}

\newcommand{\mtorBH}{m_{\rm{tor}}^{\rm{BH}}}
\newcommand{\YetorBH}{Y_{e,\rm{tor}}^{\rm{BH}}}
\newcommand{\Yetoreq}{Y_{e,\rm{tor}}^{\rm{eq}}}
\newcommand{\rtor}{r_{\rm{tor}}}
\newcommand{\mej}{m_{\rm{ej}}}
\newcommand{\Yej}{Y_{e,\rm{ej}}}
\newcommand{\vej}{v_{\rm{ej}}}
\newcommand{\XLA}{X_{\rm{LA}}}
\newcommand{\qheat}{q_{\rm{heat}}}
\newcommand{\tpm}{t_{\rm{pm}}}
\newcommand{\tmap}{t_{\rm{map}}}
\newcommand{\dd}{\mathrm{d}}
\newcommand{\cminv}{cm$^{-3}$}
\newcommand{\sinv}{s$^{-1}$}
\newcommand{\ginv}{g$^{-1}$}
\newcommand{\rhomax}{\rho_{\rm max}}

\def\ga{\,\,\raise0.14em\hbox{$>$}\kern-0.76em\lower0.28em\hbox
{$\sim$}\,\,}
\def\la{\,\,\raise0.14em\hbox{$<$}\kern-0.76em\lower0.28em\hbox
{$\sim$}\,\,}

\shorttitle{Delayed black-hole formation kilonovae}

\shortauthors{Just et al.}

\begin{document}

\title{End-to-end kilonova models of neutron-star mergers \\ with delayed black-hole formation \vspace{-1.5cm}}

\author{O.~Just\altaffilmark{1,2}, V.~Vijayan\altaffilmark{1,3}, Z.~Xiong\altaffilmark{1}, S.~Goriely\altaffilmark{4}, T.~Soultanis\altaffilmark{1}, \\ A.~Bauswein\altaffilmark{1,5}, J.~Guilet\altaffilmark{6}, H.-Th.~Janka\altaffilmark{7}, G.~Mart{\'i}nez-Pinedo\altaffilmark{1,5,8}}

\email{o.just@gsi.de}

\altaffiltext{1}{GSI Helmholtzzentrum f\"ur Schwerionenforschung, Planckstraße 1, D-64291 Darmstadt, Germany}
\altaffiltext{2}{Astrophysical Big Bang Laboratory, RIKEN Cluster for Pioneering Research, 2-1 Hirosawa, Wako, Saitama 351-0198, Japan}
\altaffiltext{3}{Department of Physics and Astronomy, Ruprecht-Karls-Universit\"at Heidelberg, Im Neuenheimer Feld 226, 69120 Heidelberg, Germany}
\altaffiltext{4}{Institut d'Astronomie et d'Astrophysique, CP-226, Universit\'e Libre de Bruxelles, 1050 Brussels, Belgium}
\altaffiltext{5}{Helmholtz Research Academy Hesse for FAIR (HFHF), GSI Helmholtz Center for Heavy Ion Research, Campus Darmstadt, Planckstra{\ss}e 1, 64291 Darmstadt, Germany}
\altaffiltext{6}{Universit\'e Paris-Saclay, Universit\'e Paris Cit\'e, CEA, CNRS, AIM, F-91191 Gif-sur-Yvette, France}
\altaffiltext{7}{Max-Planck-Institut f\"ur Astrophysik, Postfach 1317, 85741 Garching, Germany}
\altaffiltext{8}{Institut f\"ur Kernphysik (Theoriezentrum), Fachbereich Physik, Technische Universit\"at Darmstadt, Schlossgartenstraße 2, 64289 Darmstadt, Germany}

\begin{abstract}
  We investigate the nucleosynthesis and kilonova properties of binary neutron-star (NS) merger models which lead to intermediate remnant lifetimes of $\sim 0.1\text{--}1$\,seconds until black-hole (BH) formation and describe all components of material ejected during the dynamical merger phase, NS-remnant evolution, and final viscous disintegration of the BH torus after gravitational collapse. To this end we employ a combination of hydrodynamics, nucleosynthesis, and radiative-transfer tools to achieve a consistent end-to-end modeling of the system and its observables. We adopt a novel version of the Shakura-Sunyaev scheme allowing to vary the approximate turbulent viscosity inside the NS remnant independently of the surrounding disk. We find that asymmetric progenitors lead to shorter remnant lifetimes and enhanced ejecta masses, although the viscosity affects the absolute values of these characteristics. The integrated production of lanthanides and heavier elements in such binary systems is sub-solar, suggesting that the considered scenarios contribute in a sub-dominant fashion to r-process enrichment. One reason is that BH-tori formed after delayed collapse exhibit less neutron-rich conditions than typically found, and often assumed in previous BH-torus models, for early BH formation. The outflows in our models feature strong anisotropy as a result of the lanthanide-poor polar neutrino-driven wind pushing aside lanthanide-rich dynamical ejecta. Considering the complexity of the models, the estimated kilonova light curves show promising agreement with AT2017gfo after times of several days, while the remaining inconsistencies at early times could possibly be overcome in binary configurations with a more dominant neutrino-driven wind relative to the dynamical ejecta.
\end{abstract}


\section{Introduction}\label{sec:introduction}

\begin{table*}
  \centering
    \caption{\label{tab:models}Model properties and results}
    \begin{tabular}{l|ccccccccccc}
      \tableline \tableline 
      model name & mass ratio & $\nvis$ & $\avis$ & $\tcol$ & $\mtorBH$          & $\YetorBH$ & $\mej^{\rm{total}}$ & $\mej^{\rm{dyn/NS/BH}}$ & $\Yej^{\rm{dyn/NS/BH}}$ & $\vej^{\rm{dyn/NS/BH}}$ & $\XLA^{\rm{dyn/NS/BH}}$  \\
                 &            &         &         & [ms]    & [$10^{-2}\,\Msun$] &            & [$10^{-3}\Msun$]    & [$10^{-3}\Msun$]        &                         & [$10^{-2}\,c$]          & [$10^{-3}$]              \\
      \tableline                                                                                                        
      sym-n1-a6  & 1          & 1       & 0.06    & 122     & 12.5               & 0.257      & 74                  & 6/20/47                 & 0.24/0.41/0.30          & 22/18/5.5               & 142/0.00/2.85            \\
      sym-n05-a3 & 1          & 0.5     & 0.03    & 186     & 13.4               & 0.214      & 57                  & 6/21/31                 & 0.23/0.42/0.30          & 22/16/4.3               & 135/0.00/8.56            \\
      sym-n05-a6 & 1          & 0.5     & 0.06    & 104     & 14.1               & 0.255      & 76                  & 6/18/52                 & 0.24/0.42/0.31          & 22/20/5.8               & 143/0.00/2.37            \\
      sym-n10-a3 & 1          & 10      & 0.03    & 915     & 1.78               & 0.317      & 33                  & 6/21/5                  & 0.24/0.39/0.32          & 21/11/3.8               & 141/0.36/0.73            \\
      sym-n10-a6 & 1          & 10      & 0.06    & 815     & 1.87               & 0.318      & 37                  & 6/29/2                  & 0.23/0.37/0.33          & 21/10/5.1               & 147/0.52/0.00            \\
      \tableline                                                                                                        
      asy-n1-a6  & 0.75       & 1       & 0.06    & 96      & 16.2               & 0.250      & 86                  & 11/21/55                & 0.25/0.41/0.30          & 22/20/5.9               & 125/0.06/4.90            \\
      asy-n05-a3 & 0.75       & 0.5     & 0.03    & 148     & 16.3               & 0.224      & 71                  & 11/25/35                & 0.25/0.41/0.31          & 20/17/4.8               & 118/0.21/7.03            \\
      asy-n05-a6 & 0.75       & 0.5     & 0.06    & 88      & 17.5               & 0.250      & 87                  & 11/21/56                & 0.25/0.41/0.31          & 21/20/6.1               & 121/0.16/6.90            \\
      asy-n10-a3 & 0.75       & 10      & 0.03    & 680     & 5.57               & 0.252      & 61                  & 11/30/20                & 0.25/0.39/0.31          & 19/13/3.5               & 126/0.08/6.78            \\
      asy-n10-a6 & 0.75       & 10      & 0.06    & 520     & 5.77               & 0.283      & 76                  & 13/39/24                & 0.24/0.37/0.29          & 18/12/5.0               & 131/4.40/3.97            \\
      \tableline
    \end{tabular}
    \tablecomments{From left to right: mass ratio, the two parameters entering the viscosity scheme, time of BH formation, mass and average electron fraction of the torus measured at $\tcol+10\,$ms, total ejecta mass, as well as for each ejecta component (dynamical/NS-torus/BH-torus ejecta) the mass, average electron fraction (measured at temperature of $5\,$GK), average velocity, and mass fraction of lanthanides plus actinides. Values for $\XLA$ and $\qheat$ were obtained using nuclear network A.}
\end{table*}

The recent, first multi-messenger observation of a binary neutron-star (NS) merger, GW170817/AT2017gfo \citep[e.g.][]{Abbott2017b,Villar2017a,Watson2019s}, lends strong support to the idea \citep{Lattimer1977} that NS mergers are indeed significant sites of rapid neutron-capture (r-) process nucleosynthesis \citep{Arnould2007, Cowan2021g}. Simulations of NS mergers and their aftermath predict that r-process viable outflows can be produced in each of the following three phases: during and right after the merger (called dynamical ejecta; e.g. \citealp{Goriely2011,Korobkin2012,Wanajo2014a}), from a NS-torus remnant in the case that it is formed \citep[e.g.][]{Perego2014a,Metzger2014,Fujibayashi2018a,Mosta2020c}, and from a black-hole (BH) torus remnant formed promptly or after collapse of a NS remnant \citep[e.g.][]{Fernandez2013b,Just2015a,Siegel2018c,Fujibayashi2020a}. Depending on the masses of the initial two NSs and the nuclear equation of state (EOS), those components can make different contributions to and, thus, have different relative importance for the nucleosynthesis yields and the electromagnetic kilonova (KN) counterpart.

The role of each component is not well constrained so far, neither theoretically nor observationally (i.e. based on AT2017gfo). One reason is that most previous theoretical studies treat each component individually or separately, and so far only a few studies discuss models with a consistent inclusion of all components. \citet{Fujibayashi2020b} and \citet{Shibata2021c}, in combination with the corresponding KN studies of \citet{Kawaguchi2021a, Kawaguchi2022o}, reported models with a very long-lived NS remnant, which do not produce ejecta from a BH-torus system. On the other hand, \citet{Fujibayashi2023a} considered systems in which matter ejection from the NS remnant is terminated early on, because the remnant collapses just shortly after the merger. Recently, \citet{Kiuchi2022a} reported a magneto-hydrodynamic (MHD) simulation covering the first second of evolution of a similarly short-lived case, which confirmed the basic results by \citet{Fujibayashi2023a} obtained using a more approximate $\alpha$-viscosity scheme \citep{Shakura1973}.

In this Letter, we present the first end-to-end models of NS mergers with intermediate remnant lifetimes (between $\sim \! 0.1\text{--}1\,$s). These systems are distinguished from the aforementioned scenarios as they yield roughly comparable amounts of all three types of ejecta. Different from previous long-term evolution models of NS remnants \citep[e.g.][]{Perego2014a,Metzger2014,Fujibayashi2020a}, our simulations adopt an energy-dependent neutrino transport scheme as well as an improved $\alpha$-viscosity scheme guided by MHD results.

These first neutrino-viscous models of mergers with significantly delayed BH formation lead to several new insights: 1) The lifetime of the NS remnant in such type of systems is shorter for asymmetric than for symmetric binaries, and it depends sensitively on the viscosity inside the NS. 2) Ejecta launched during the BH-torus phase are less neutron rich than predicted by models using manually-constructed initial conditions. 3) In the considered systems of intermediate lifetimes, the synthesis of lanthanides and heavier elements is not efficient enough to explain the solar pattern. 4) The combination of all ejecta components is significantly more anisotropic than just the dynamical ejecta because of a massive, dominantly polar, neutrino-driven outflow from the NS remnant. 5) The KN produced by the combined ejecta can (may not) shine bright enough to explain AT2017gfo at late (early) times. 6) For a given viscosity both the summed mass of all ejecta components, as well as their individual contributions, are systematically higher for asymmetric than for equal-mass binaries.

After outlining our model setup in Sect.~\ref{sec:model-setup}, we will report on the aforementioned findings in Sect.~\ref{sec:results}, and discuss some implications in Sect.~\ref{sec:discussion}. The Appendices provide additional information regarding selected properties of our models.

\begin{figure*}
  \includegraphics[trim=10 12 5 5,clip,width=0.99\textwidth]{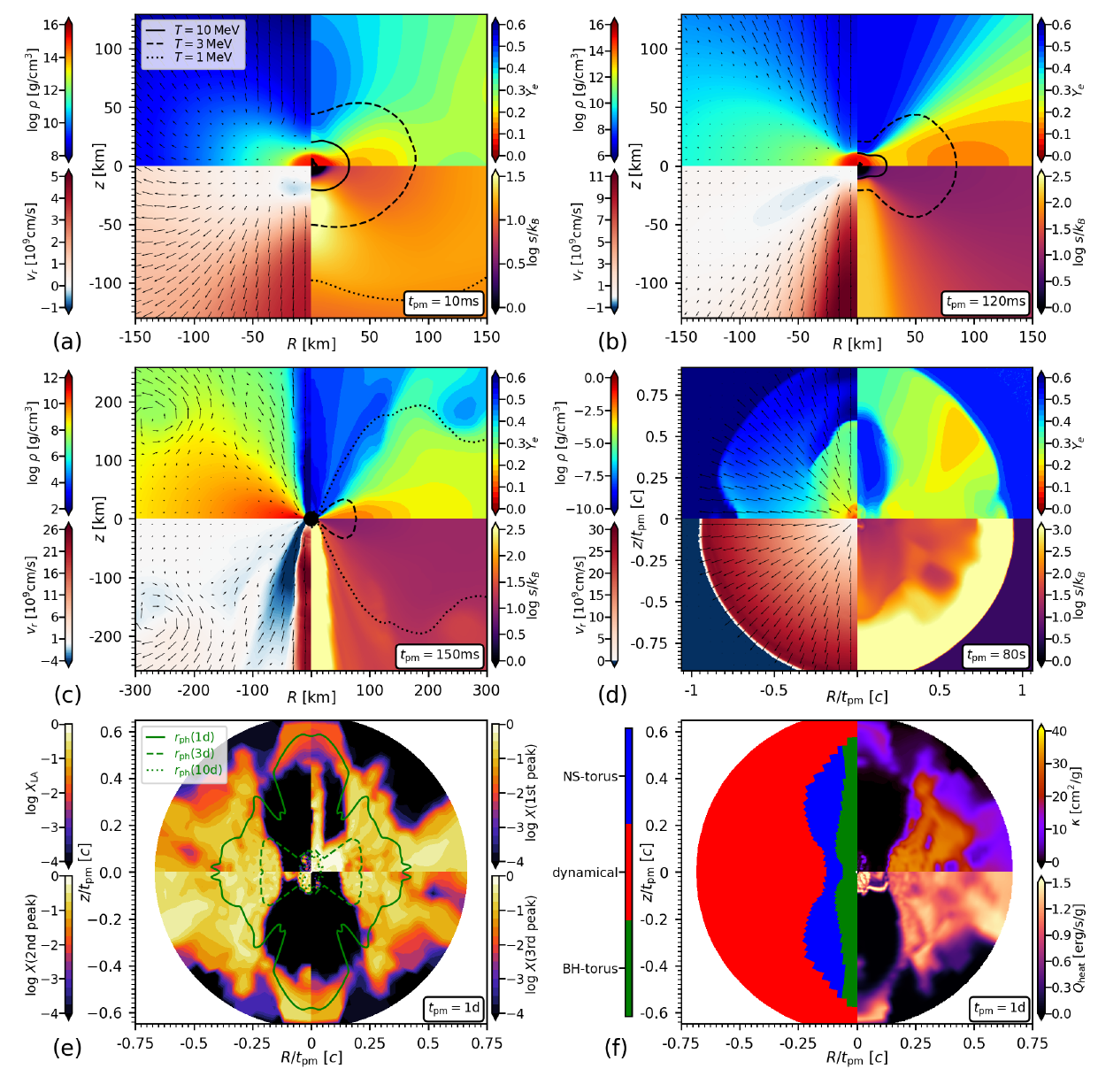}
  \caption{Snapshots of model sym-n1-a6 at different post-merger times, $\tpm$. Panels~(a)-(d) show the density $\rho$, radial velocity $v_r$, electron fraction $Y_e$, and entropy per baryon $s$, as well as velocity arrows (left sides) and contours of temperature $T$ (right sides). Panel~(e) shows mass fractions of lanthanides plus actinides $\XLA$, and mass fractions of elements in the 1st, 2nd, and 3rd r-process peak, overlaid with green lines denoting the time-dependent location of the radial photosphere (computed as in \citealp{Just2022a}). Panel~(f) shows a map color-coding the three main ejecta components, the opacity $\kappa$, and the effective radioactive heating rate $Q_{\rm heat}$. Panels~(a)-(d) show data from both hemispheres and panels~(e),~(f) from just the northern hemisphere assuming equatorial symmetry.}
  \label{fig:cont}
\end{figure*}

\section{Model setup}\label{sec:model-setup}

Each model consists of three successive hydrodynamics simulations and two post-processing steps that provide the nucleosynthesis yields and KN light curve. The hydrodynamical evolution of the merger is followed with a 3D general relativistic (GR) smoothed-particle hydrodynamics (SPH) code \citep{Oechslin2007,Bauswein2010c} that employs a modern leakage-plus-absorption scheme (ILEAS; \citealp{Ardevol-Pulpillo2019a}) to describe neutrino cooling and heating, including electron-neutrinos ($\nu_e$), electron-antineutrinos ($\bar\nu_e$), and a third species ($\nu_x$) representative of all heavy-lepton neutrinos. At a post-merger time, $\tpm$, of $\tpm=\tmap=10\,$ms we azimuthally average the SPH configuration, map it to a spherical polar grid, and, assuming axisymmetry, continue the post-merger evolution using the special-relativistic code ALCAR-AENUS \citep{Obergaulinger2008a, Just2015b}, which adopts an energy-dependent M1 neutrino-transport scheme. We employ the same general relativistic corrections in the transport equations and the same neutrino-interaction rates and formulations that have been used in \citet{Just2018b} (and are based on \citealp{Bruenn1985,Hannestad1998,Horowitz2002a,Pons2000}), except that in the present study we neglect inelastic neutrino-electron scattering and use the approximation by \citet{OConnor2015a} to describe pair processes. For the transition from the SPH simulations (which do not evolve local neutrino energy- and flux-densities), the neutrino energies are initially (i.e. at $\tpm=\tmap$) set to Fermi-distributions corresponding to the local thermodynamic state above densities of $10^9\,$g\,\cminv and vanish everywhere else, and the fluxes vanish everywhere.

The numerical settings adopted in the SPH simulations are the same as detailed in \citet{Ardevol-Pulpillo2019a,Kullmann2021a}. The total number of SPH particles is $3\times 10^5$, and the neutrino source terms are computed on a uniform cartesian grid having 305 cells of size $738\,$m along each direction (see Appendix~\ref{sec:resol-depend-sph} for a test of the neutrino-grid resolution; uncertainties in the dynamical ejecta masses due to the particle resolution are estimated to be several ten percent \citep{Bauswein2013} comparable to estimated error bars of grid-based merger simulations \citep[e.g.][]{Radice2018b}). The post-merger simulations are conducted using a radial grid with constant cell size of $\Delta r=100\,$m within radii of $r<20\,$km and afterwards increasing by $\approx 2.3\,\%$ per cell, and a uniform polar grid with a resolution of $2.25^\circ$. The neutrino-energy range between 0 and 400\,MeV is discretized using 15 bins, of which the size increases by 40\,\% per bin. In order to prevent extremely small time-integration steps, we assume at radii below $1.5\,$km a uniformly rotating core with spherically symmetric thermodynamic properties. We verified that post-merger models with higher resolution and smaller 1D core produce essentially the same results.

In order to reduce the inconsistency between the curved-spacetime merger models and flat-spacetime post-merger models, we map the primitive variables such that the radial volume element of the SPH model $\psi^6\dd(r_{\rm{SPH}}^3/3)$ (with conformal factor $\psi$; cf. \citealp{Oechslin2007}) equals the post-merger volume element $\dd(r^3/3)$ along each radial direction -- thus approximately preserving volume integrals of conserved variables (baryonic rest mass etc.) -- and we define the three-velocities in the post-merger model as functions of the corresponding SPH velocities \citep[see][]{Oechslin2007} as $v^i=v^i_{\rm SPH}\psi^2$. Gravitation is treated by solving a Poisson equation augmented with relativistic corrections \citep{Muller2008}, in which the monopole contribution is replaced either by an effective TOV potential \citep{Marek2006} (for times $\tpm$ earlier than the time of BH formation, $\tcol$) or by the pseudo-Newtonian BH potential of \citet{Artemova1996} (for $\tpm>\tcol$). Once the NS remnant becomes gravitationally unstable (i.e. at $\tpm=\tcol$) we replace the innermost region by an outflow boundary mimicking the central BH, while consistently updating its size, mass, and angular momentum through time integration of the boundary fluxes.

For describing turbulent viscosity driven by the magneto-rotational instability (MRI), we extend the classical $\alpha$-viscosity scheme by \citet{Shakura1973} such that it can capture MRI-related viscosity in both the rotation-supported regime (i.e. the accretion torus) and the pressure-supported regime (i.e. the NS remnant). In a pressure-supported object with a subsonic shear velocity, the MRI-driven viscosity is indeed not expected to scale with the sound speed but rather to behave in a quasi-incompressible manner \citep{Reboul-Salze2021z,Reboul-Salze2022a}. Our formulation therefore expresses the kinematic viscosity as:
\begin{equation}\label{eq:etavis}
  \nu_{\rm vis} = \avis H_{\rm vis}^2\, |\Omega|\, \tilde{q}^{\nvis} \, ,
\end{equation}
namely the product of the generalized characteristic length scale
\begin{equation}\label{eq:hvis}
  \Hvis = \min\{\left|\rho/\nabla\rho\right|,r,c_i/\Omega_{\rm K}\}
\end{equation}
(with density $\rho$, spherical radius $r$, isothermal sound speed $c_i=\sqrt{P/\rho}$, gas pressure $P$, and Keplerian angular velocity $\Omega_{\rm K}$) and the characteristic velocity scale $\Hvis\Omega$ (with angular velocity $\Omega$). The additional quenching factor
\begin{equation}\label{eq:qtil}
  \tilde{q}^{\nvis} = \begin{cases}
    0 \, , & \text{if } \dd \Omega/\dd R > 0 \, , \\
    1 \, , & \text{else if } \left|\dd \ln\Omega/\dd \ln R \right| > q_0 \, ,\\
    \left(\frac{1}{q_0}\left|\frac{\dd \ln\Omega}{\dd\ln R}\right| \right)^{\nvis} \, , & \text{else}
    \end{cases}
\end{equation}
(with cylindrical radius $R$) accounts for the tendency \citep[e.g.][]{Pessah2008h} of the MRI to be reduced in regions where the shear is sub-Keplerian, i.e. $q=\left|\dd\ln\Omega/\dd\ln R\right|<q_0\sim 1.5$\footnote{We here choose $q_0=1.7$ to be slightly higher than the Newtonian value of $1.5$ because of our steeper-than-Newtonian gravitational potential.}. The parameter $\nvis$ thus varies the strength of turbulent viscosity in the NS remnant (where $q<q_0$) relatively independently of that in the surrounding disk (where $q\simeq q_0$). This allows to explore parametrically the sensitivity to the viscosity inside the NS remnant (which is poorly constrained so far by existing simulations; \citealp{Kiuchi2018a,Palenzuela2022a,Margalit2022a}), while keeping the viscosity in the disk (which is known to be fairly well reproduced by a conventional $\alpha$-viscosity scheme; \citealp{Fernandez2019b,Just2022b,Hayashi2021a}) unchanged. Significant uncertainty also comes from the dependence on the diffusive processes through the magnetic Prandtl number \citep{Guilet2022a,Held2022a}, which justifies exploring different parameter values.

At $\tpm=10\,$s the inner (outer) radial boundary is moved to a radius of $10^4\,$km ($4\times10^7\,$km), and a third simulation is conducted to follow the expansion of just the ejected material until $\tpm=100\,$s. The ejecta configuration at $\tpm=100\,$s is assumed to be homologous, with $r(\tpm)=v\,\tpm$, and equatorially symmetric and gets sampled in the northern hemisphere up to velocities of 0.7\,$c$ by 1000\textit{--}2000 tracer particles per model, the time-evolution of which is obtained by path-integration backward in time using the available simulation outputs. The sampling of the dynamical ejecta takes into account the entire evolution, i.e. utilizes data also from the SPH simulations by splitting trajectories constructed from post-merger simulation data at $\tpm=\tmap$ and associating them with a number of SPH particles. Since this step, by which the effective number of tracers is increased to 4000\textit{--}5000 per model, is non-trivial, we provide further details on this procedure in Appendix~\ref{sec:mapp-dynam-ejecta}. The tracers are input to a nuclear network solver that predicts the nucleosynthesis yields. We use two independent solvers here (called network A and B hereafter), allowing us to cross-validate the yields and heating rates and to isolate uncertainties related to the network code and its physics input from other modeling uncertainties. Network A (used previously in, e.g., \citealp{Goriely2011,Just2015a,Kullmann2022a}) takes nuclear ingredients from experiments where available and, where not, from theoretical models, namely nuclear masses from the BSkG2 mass model \citep{Ryssens2022a}, $\beta$-decay rates from \citet{Marketin2016a}, reaction rates from TALYS estimates with microscopic inputs \citep{Goriely2018b}, including BSkG2 masses, and fission probabilities and fragment distributions from \citet{Lemaitre2021a}. Network B (employed previously in, e.g., \citealp{Wu2016a,Collins2023a}) uses the reaction rates for neutron captures, photo-dissociation and fission based on the HFB21 mass model \citep{Goriely2010} as described in \citet{Mendoza2015} and $\beta$-decay rates from \citet{Marketin2016a}.

Finally, for assessing the KN light curve the tracers, including their composition and radioactive heating rates, are used as input for an approximate photon transport scheme to estimate the KN light curve in the same way as detailed in \cite{Just2022a}.

Table~\ref{tab:models} summarizes the parameters for all investigated models. We consider both a symmetric (ratio of gravitational masses of $M_1/M_2=1$) and asymmetric ($M_1/M_2=0.75$) progenitor configuration (with $M_1+M_2=2.75\,\Msun$) and for both cases vary the viscosity parameters $\nvis\in\{0.5,1,10\}$ and $\avis\in\{0.03,0.06\}$. The SFHo EOS \citep{Steiner2013}, extended to low densities with a four-species EOS \citep[e.g.][]{Just2015a}, is adopted.\footnote{We note that some GR merger studies \citep[e.g.][]{Radice2018b,Fujibayashi2023a} report early ($\tpm< 20\,$ms) BH formation for a similar total binary mass and EOS used here, indicating that our post-merger gravity treatment may be slightly weaker than a GR treatment. However, discrepancies concerning the collapse behavior, i.e. the threshold mass for prompt BH formation, also exist between full-GR simulations \citep[e.g.][]{Kolsch2022a}. Since the remnant lifetime is expected to be very sensitive to the total mass, large differences in the remnant lifetime effectively correspond to small discrepancies in the total binary mass, which is why we anticipate that our calculations reliably capture the scenario of a merger remnant with intermediate lifetime.}

\begin{figure*}
  \includegraphics[width=0.99\textwidth]{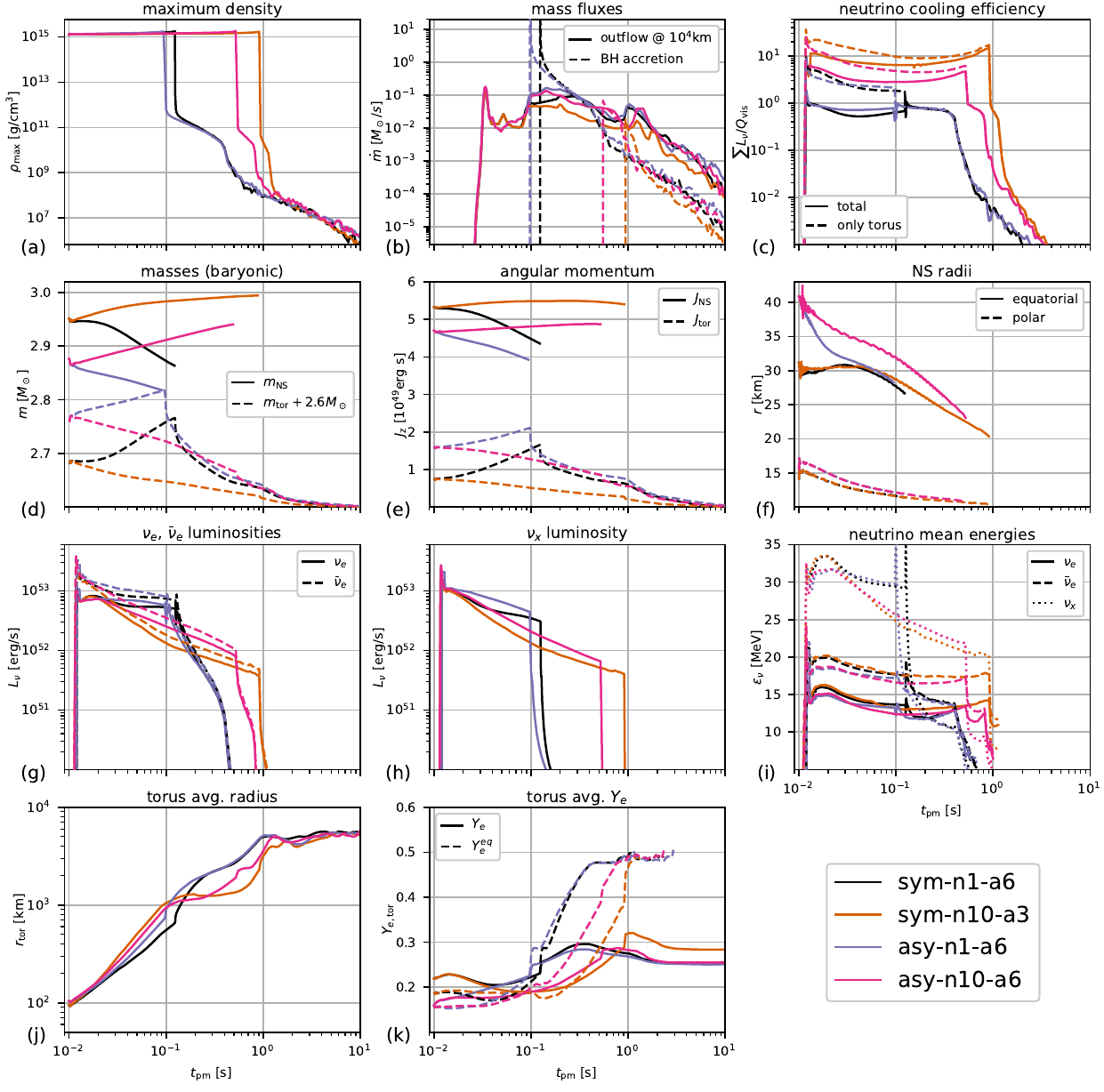}
  \caption{Global properties of the four models mentioned in the bottom right, namely the maximum density (panel~(a)), outflow mass fluxes through the sphere at $r=10^4\,$km and mass fluxes into the central BH once formed (panel~(b)), ratio of the total neutrino luminosities to the volume-integrated viscous heating rate (panel~(c)), masses of NS and torus (panel~(d)), angular momenta of NS and torus (panel~(e)), radii of the NS surface in equatorial and polar direction (panel~(f)), luminosities of electron-type neutrinos (panel~(g)) and heavy-lepton neutrinos (panel~(h)), neutrino mean energies (computed as the ratio of energy-to-number fluxes; panel~(i)), mass-averaged radius of the torus ($\int r \dd m /\int\dd m$; panel~(j)), and mass-average of the torus electron fraction ($\int Y_e \dd m /\int\dd m$) and its equilibrium value $Y_e^{\rm eq}$ (panel~(k)). The torus is defined as all material below $r=10^4\,$km having $\rho<10^{12}\,$g\,\cminv and the NS as all material with $\rho>10^{12}\,$g\,\cminv. All neutrino-related quantities are measured in the lab-frame at $r=500\,$km by an observer at infinity. The neutrino fluxes vanish initially (at $\tpm\la \tmap= 10\,$ms) because the plot shows only data from the post-merger simulations (which are initialized at $\tmap$ with vanishing neutrino fluxes).}
  \label{fig:glob}
\end{figure*}

\section{Results}\label{sec:results}

The following sections address the collapse behavior, torus properties, nucleosynthesis yields, ejecta geometry, and KN signal. Figure~\ref{fig:cont} illustrates snapshots at different times for model sym-n1-a6, and Fig.~\ref{fig:glob} shows the time evolution of global properties for several models. The $Y_e$ distribution, nucleosynthesis yields, and radioactive heating rate are depicted in Fig.~\ref{fig:abund}, and KN observables are provided in Fig.~\ref{fig:kilo}. We adopt the (somewhat ambiguous) criterion $\rho>10^{12}\,$g\,cm$^{-3}$ to discriminate material located in the NS remnant from that in the surrounding torus.

\subsection{Lifetime dependence on mass ratio and viscosity}\label{sec:life-time-dependence}

Even though our post-merger models are not performed in GR, the adopted TOV potential is known to compare well with GR solutions (at least in the case of core-collapse supernovae; \citealp{Liebendorfer2005}) and, importantly, it captures the existence of a maximum mass above which the configuration becomes gravitationally unstable \citep[][]{Marek2006,Muller2008}. Our models leading to meta-stable NSs thus allow to obtain a first, basic idea of the way how spectral neutrino transport and viscosity together act in driving the remnant towards instability, in dependence on the mass ratio and the chosen strength of the viscosity.

Shortly after the merger a pressure-supported, nearly uniformly-rotating NS core is formed, surrounded by a rotation-supported, nearly Keplerian-rotating torus (cf. Fig.~\ref{fig:angvel}). Both angular momentum transport and neutrino cooling cause a continuous growth of the maximum density, $\rhomax$ (cf. panel~(a) of Fig.~\ref{fig:glob}) until eventually the NS becomes gravitationally unstable and forms a BH. We find the BH-formation times, $\tcol$ (cf. Table~\ref{tab:models}), to be systematically shorter (by $\sim 20\text{--}40\,\%$) in the asymmetric compared to the symmetric models for a given viscosity. This result is likely to be a consequence of the tendency that in our asymmetric merger models a relatively large fraction of angular momentum ends up in the torus, as opposed to the NS, at $\tpm=\tmap$, resulting in NS remnants that rotate with a smaller (both absolute and mass-specific) angular momentum compared to the symmetric case (cf. panels~(d)~and~(e) of Fig.~\ref{fig:glob}). This tendency appears plausible from the point of view of Newtonian point-particle dynamics: In an asymmetric binary the low-mass star revolves around the center of mass (COM) at a wider orbit and, therefore, with higher angular momentum than a star in a symmetric binary with the same orbital separation \citep[see, e.g.,][]{Bauswein2021a}. As a result, once tidal effects disrupt the low-mass star, relatively more high-angular-momentum material is located further away from the COM and more efficiently transferred from the high-density NS remnant into the surrounding torus. We checked that this tendency also appears for a different EOS and using an entirely different hydrodynamics code (cf. Appendix~\ref{sec:mass-ratio-depend}), however, future investigations will need to further scrutinize this tendency as well as its consequence for the remnant lifetime.

We also observe a mild increase of the ejecta mass for asymmetric compared to equal-mass systems, both for the sum of all ejecta as well as for each component individually (cf. $\mej$ in Table~\ref{tab:models}).

Our models, however, also show that different viscosities can alter $\tcol$ and $\mej$ even more dramatically than the mass ratio, implying that a solid understanding of the NS viscosity is required to firmly connect the remnant lifetime and binary properties. We find shorter lifetimes for higher viscosities inside the NS. This suggests that with increasing viscosity the loss of angular momentum (pushing the NS towards instability; cf. panel~(e) of Fig.~\ref{fig:glob}) has a stronger impact than the loss of mass (that tends to stabilize the NS; cf. panel~(d) of Fig.~\ref{fig:glob}), at least in our quasi-Newtonian post-merger models. Future GR models will have to check the robustness of this finding.

\subsection{Torus properties at black-hole formation}\label{sec:torus-neutr-richn}

The properties of the torus at the time of BH formation are important parameters determining the nucleosynthesis signature of BH-torus systems formed after mergers. Existing compilations of the torus mass, $\mtorBH$, for a given NS binary and EOS \citep[e.g.][]{Kruger2020a} are, however, based on simulations covering only the merger, but not the post-merger evolution, and therefore cannot accurately predict torus properties in the case of late-time ($\tpm\ga 20\,$ms) BH-torus formation. 

In our models that account for the neutrino-viscous evolution of the NS remnant until BH-formation, we find that the torus mass can both grow or decrease during the NS-remnant evolution (cf. panel~(d) of Fig.~\ref{fig:glob}), hence causing a significant variation of $\mtorBH$ (cf. Table~\ref{tab:models}) between models with different viscosities. This behavior is mainly a result of the competition between viscous angular momentum transport in the NS remnant (which tends to push material radially outward) and in the torus (which drives accretion onto the NS). For high NS viscosity (i.e. low values of $\nvis$), angular momentum is transported by the NS faster than by the disk, hence causing a net loss of mass and angular momentum of the NS; cf. models with $\nvis=1$ in panels~(d)~and~(e) of Fig.~\ref{fig:glob}. The opposite tendency is observed for models with $\nvis=10$. The two competing effects are additionally superimposed by neutrino cooling, which gradually makes the entire configuration more compact (cf. panel~(f) of Fig.~\ref{fig:glob}) and thereby tends to increase (decrease) the NS (torus) mass.

Other important parameters of the torus, apart from its mass, are its $Y_e$ and radial size. Many existing studies \citep[e.g.][]{Fernandez2013b,Just2015a,Siegel2018c} take manually-constructed equilibrium tori as initial conditions, and assume those to be neutron-rich ($Y_e\approx 0.1$) and rather compact (with mass-averaged radii of $r_{\rm tor}\approx 50\text{--}100\,$km). In our models, the torus undergoes significant viscous spreading already before BH formation, causing its mass-averaged radius to grow substantially until $\tpm=\tcol$, by factors of $\sim 5\text{--}20$ compared to the initial value of $\rtor(\tpm=\tmap)\approx 100\,$km (cf. panel~(j) of Fig.~\ref{fig:glob}). This viscous pre-evolution of the torus results in the electron degeneracy to be lower and, therefore, the equilibrium electron fraction, $\Yetoreq$ (computed using Eq.~(3) of \citealp{Just2022b}), to be higher at the time of BH-torus birth, $\tcol$, compared to the early values at $\tmap=10\,$ms (cf. panel~(k) of Fig.~\ref{fig:glob}). The values of $Y_e$ in the torus at $\tpm=\tcol$ are therefore relatively high, $\YetorBH\approx$0.22-0.32 (cf. Table~\ref{tab:models}), which has important consequences for the nucleosynthesis signature of the BH-torus outflows (cf. Sect.~\ref{sec:nucl-yields}).

\subsection{Ejecta interaction and geometry}\label{sec:ejecta-inter-geom}

\begin{figure*}



  \includegraphics[width=\textwidth]{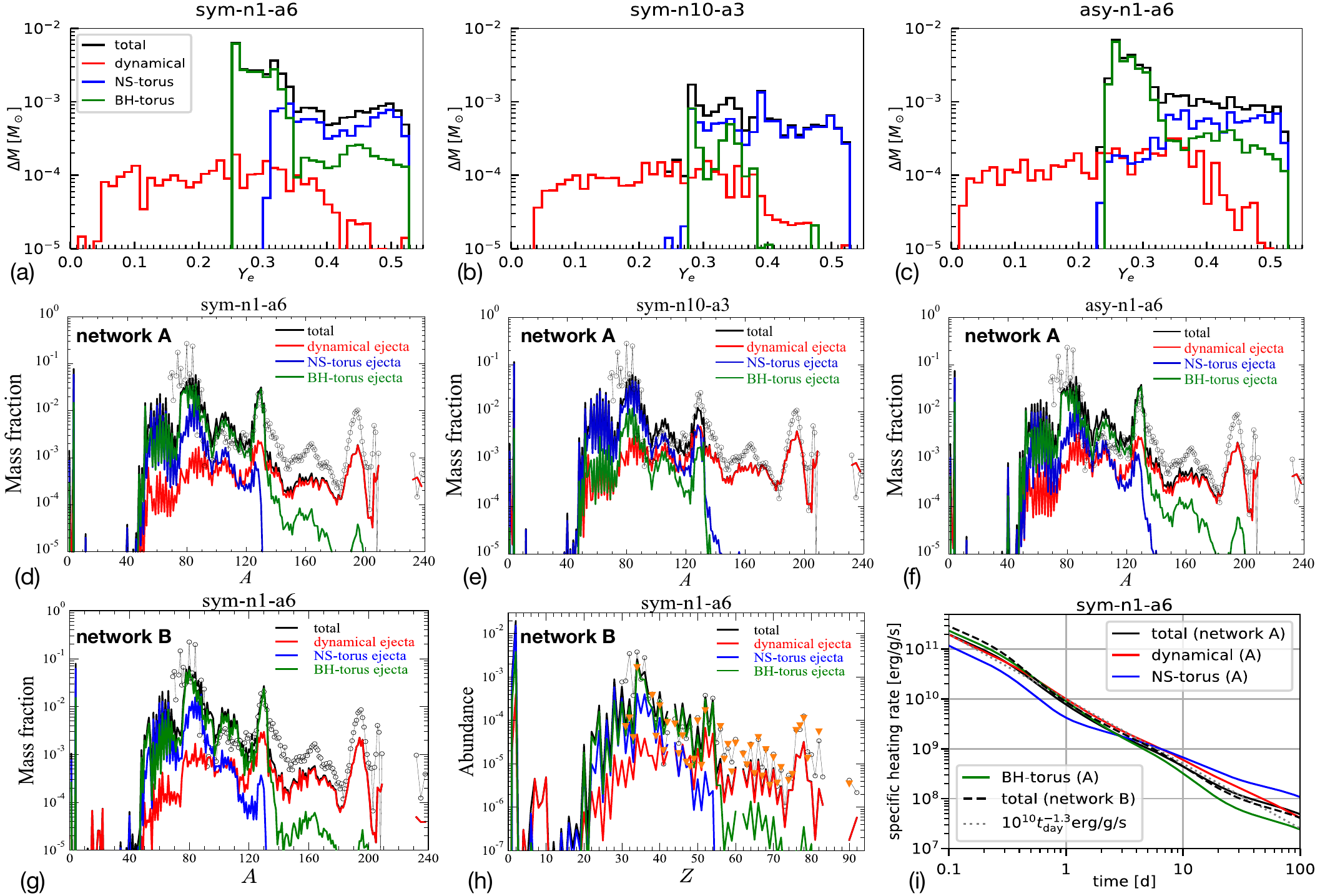}

  \caption{Mass versus $Y_e$ histograms for models sym-n1-a6, sym-n10-a3, and asy-n1-a6 (panels~(a)-(c)) and the corresponding mass fractions of synthesized elements versus atomic mass number using nuclear network A (panels~(d)-(f)) for each ejecta component and for the total ejecta (see labels). The third row shows for model sym-n1-a6 yields versus atomic mass number (panel~(g)) and elemental abundances (panel~(h)) obtained with network B, as well as the specific radioactive heating rate for the indicated ejecta components and networks. All yields are shown for a time (typically about 100\,Myr) when all elements, except the three longest-lived Th and U isotopes, have decayed into stable nuclei. Black circles in panels~(d)-(h) show solar r-process yields \citep{Goriely1999a}, scaled to the predicted total yields of Sr, the only confirmed element in AT2017gfo \citep{Watson2019s,Domoto2021a,Gillanders2022a}. Orange triangles in panel~(h) denote abundances observed for the metal-poor star HD-222925 \citep{Roederer2022b} scaled to match the solar Eu abundance. The grey, dotted line in panel~(i) shows for reference the heating rate $10^{10}\times(\tpm/ 1\,\mathrm{d})^{-1.3}$\,erg\,\ginv\,\sinv}.
  \label{fig:abund}
\end{figure*}

In order to deduce from an observed KN the mass and other properties of the ejected material, theoretical models must be able to predict the final, spatial distribution of the total ejecta, a task that can only be accomplished by end-to-end models that capture the launch and expansion of all ejecta components and their dynamical interaction with each other.

The dynamical ejecta, defined here as all\footnote{Note that we do not need to impose an additional criterion to filter out gravitationally bound from unbound material, because the time at which we identify ejecta (100\,s) is late enough to ensure that all material counted as ejecta is indeed gravitationally unbound.} material fulfilling $r(\tmap) > 250\,$km, are launched during the merger in a roughly spherical fashion \citep{Bauswein2013,Hotokezaka2013b}. During the subsequent evolution of the NS remnant ($\tmap<\tpm<\tcol$) neutrino emission, starting off at rates of $\sim 10^{53}\,$erg\,\sinv per neutrino species and mean energies of $\sim$ 15, 20,~and~30\,MeV for $\nu_e$, $\bar\nu_e$,~and~$\nu_x$, respectively (cf. panels~(g)-(i) of Fig.~\ref{fig:glob}), drives a thermal wind from the NS surface with half-opening angle of $\sim 20\text{--}40^\circ$ towards both polar directions. This neutrino-driven wind (NDW), which in most of our models dominates matter ejection during the NS-torus phase, drills through large parts (up to velocities of $v/c\sim 0.5\text{--}0.6$) of the dynamical ejecta, pushing most of them away from the rotation axis while accelerating near-axis material in front of the NDW. By doing so, the NDW strongly enhances the anisotropy of the ejecta compared to that of the original dynamical ejecta (see panels~(a)~and~(b) of Fig.~\ref{fig:kilo}, or compare contours of $Y_e$ and $\kappa$ between Fig.~\ref{fig:cont} of the present study and Fig.~2 of \citealp{Just2022a}). We note that \citet[][]{Fujibayashi2020a, Kawaguchi2021a, Kawaguchi2022o} report a similar anisotropy for their models of long-lived NS remnants.

While the velocities in the NDW are spread between $0.05\la v/c\la 0.6$, the average velocity lies at about $v/c\sim 0.2$ in most models (cf. Table~\ref{tab:models}). This value is significantly higher than corresponding values reported in studies using a more approximate description of neutrino effects and of the central NS \citep[][]{Dessart2009,Perego2014a,Fahlman2018a}, though seemingly well in agreement with \citet[][]{Fujibayashi2020a}, who adopt a grey leakage-plus-M1 scheme in GR. We demonstrate in Appendix~\ref{sec:relat-import-neutr} that this fast polar outflow is indeed driven by neutrino heating, mostly by neutrino captures on free nucleons but with an additional boost due to neutrino pair annihilation. Given the intrinsic angular structure of the NDW, with the highest velocities being reached close to the polar axis, it can be assumed that multi-dimensional effects, such as collimation by the other ejecta components, play a relevant role for explaining the high velocities. For stronger viscosity in the NS remnant (i.e. lower $\nvis$ or higher $\avis$) the neutrino luminosities, and therefore the NDW mass fluxes (cf. panel~(b) of Fig.~\ref{fig:glob}), are higher at given times due to faster dissipation of rotational kinetic into thermal energy. However, due to the reduced NS lifetimes, the total mass of the NS-torus ejecta (counted here as all material not being dynamical ejecta and fulfilling $r(\tcol)>1000\,$km) shows only a modest sensitivity to viscosity, $\mej^{\rm NS}\approx 0.02\text{--}0.04\,\Msun$ (cf. Table~\ref{tab:models}), in particular more modest than in the models of long-lived NSs reported by \citet[][]{Fujibayashi2020a}, in which the ejecta from the torus (which tends to be more massive for higher viscosity) are launched entirely during the lifetime of the NS remnant.

Once the NS collapses, the neutrino luminosities quickly decrease and NDWs are mostly shut off. Consistent with previous studies using viscous equilibrium BH-tori \citep{Fernandez2013b, Just2015a, Fujibayashi2020a}, viscous matter ejection becomes operative once neutrino cooling starts to become inefficient and dominated by viscous heating (cf. panel~(c) of Fig.~\ref{fig:glob}), and it produces an outflow of roughly spherical geometry (cf. reddish region in density map in panel~(d) of Fig.~\ref{fig:cont}). This viscously-driven (and dominant) part of the BH-torus outflow carries away about $20\text{--}40\,\%$ of the torus mass at BH formation, $\mtorBH$, i.e. it inherits the uncertainties connected to viscosity imprinted on $\mtorBH$ (cf. Sec.~\ref{sec:torus-neutr-richn}). Due to their low velocities of $\vej^{\rm BH}\sim$0.03-0.06$\,c$ (cf. Table~\ref{tab:models}), the viscous BH-torus ejecta barely interact with the faster outflow components ejected earlier.

In models with high values of $\mtorBH$ we also observe, similarly as in \cite{Just2016}, an additional BH-torus outflow component, namely a jet-like outflow powered by neutrino-antineutrino pair annihilation, which transports a small amount of torus material in a narrow stream along the rotation axis (cf. Fig.~\ref{fig:cont}), reaching up to velocities of $0.5\text{--}0.6\,c$ but being unable to break out from the dynamical ejecta owing to insufficient energy supply\footnote{A more powerful jet that is able to break out (such as observed with GW170817; \citealp{Mooley2018t}) may be powered through the general relativistic Blandford-Znajek process (\citealp{Blandford1977}; see, e.g., \citealp{Gottlieb2022a} for recent numerical models), which our post-merger simulations are unable to describe.}. However, due to its low mass and relatively small volume, this choked jet has only a very small impact on the overall nucleosynthesis pattern and kilonova signal.

\subsection{Nucleosynthesis yields}\label{sec:nucl-yields}

In all our models the dynamical ejecta (Fig.~\ref{fig:abund}, red lines) are the main source of material with $Y_e< 0.25$ and $A> 140$, despite having a subdominant mass among the three ejecta components (cf. Table~\ref{tab:models}). We find their nucleosynthesis patterns to be very similar to those reported in previous studies neglecting the post-merger evolution \cite[e.g.][]{Kullmann2021a}, which suggests that the long-term evolution and dynamical interaction with other ejecta components has only a small impact on the nucleosynthesis pattern. In particular, the dynamical-ejecta yields are found to be nearly insensitive to the adopted viscosity parameters (cf. Fig.~\ref{fig:dynmap})\footnote{We do find, however, noticeable (though small) model-to-model variations in the $Y_e$ histograms, which may partially be attributed to discretization errors introduced by the limited number of tracers used to sample the dynamical ejecta (see Appendix~\ref{sec:mapp-dynam-ejecta}). These errors may also explain why our $\Yej^{\rm dyn}$ values tend to be slightly lower in the symmetric than in the asymmetric models, while the original SPH data shows the opposite tendency. Nevertheless, given the good agreement of the abundance patterns, we deem these errors to be small enough to not affect the conclusions of our study.}.

Both the NS- and BH-torus ejecta (blue and green lines in Fig.~\ref{fig:abund}, respectively) exhibit $Y_e$ values distributed broadly between $\approx 0.25\text{--}0.5$ with little amounts, if any, of material having $Y_e<0.25$. For the NDW-dominated NS-torus ejecta, high values of $Y_e$ are expected, because the equilibrium $Y_e$ in NDWs is mainly determined by neutrino absorption \citep{Qian1996}, and similar results have been reported also by studies using simpler neutrino treatments \citep[e.g.][]{Perego2014a,Metzger2014,Fujibayashi2018a}. The BH-torus ejecta, however, are less neutron rich than predicted by previous models based on manually-constructed equilibrium tori and with similar viscosity treatment \citep[e.g.][]{Fernandez2013b, Just2015a, Wu2016a}. The reason for this difference is the viscous evolution of the torus before BH formation (cf. Sect.~\ref{sec:torus-neutr-richn}) that leads to less neutron-rich and less compact tori than assumed in those previous studies. Both conditions are detrimental for the production of neutron-rich ejecta as discussed in, e.g., \cite{Fernandez2020a, Just2022b,Haddadi2023a}. The nucleosynthesis patterns of both post-merger ejecta components are, complementary to the dynamical ejecta, mainly composed of light ($A\la 140$) r-process elements, including $^{88}_{38}$Sr, but also significant amounts of iron-group elements and $^4_{2}$He (see panel~(h) of Fig.~\ref{fig:abund} for a plot showing the elemental abundances for model sym-n1-a6).

The combined yield distribution is rather insensitive to the viscosity and binary mass ratio. In all models it resembles the solar r-process pattern in the $A\la 140$ domain while falling short of heavier elements by factors of a few compared to solar. Remarkably, a smaller torus mass, $\mtorBH$, (hence a smaller amount of BH-torus ejecta) such as resulting in the symmetric model with low viscosity, sym-n10-a3 (cf. panel~(e) of Fig.~\ref{fig:abund}), leads to better agreement with the solar distribution among our models. This is because in those models the relative contribution from dynamical ejecta is greater and, consequently, the ratio of $A<140$ to $A>140$ yields is reduced compared to models exhibiting higher values of $\mtorBH$.

In panel~(h) of Fig.~\ref{fig:abund} we also compare to the abundance pattern measured in the metal-poor star HD-222925 \citep{Roederer2022b}, which provides a nearly complete r-process stellar abundance. The abundance pattern of light (1st- and 2nd-peak) r-process elements, which has been considered a challenge for nucleosynthesis models \citep{Holmbeck2022a}, is reproduced remarkably well.

The nuclear heating rate per mass unit is a crucial quantity determining the KN signal. As shown in panel~(i) of Fig.~\ref{fig:abund} for model sym-n1-a6, the heating rate before thermalization (i.e. reduced by neutrino contributions but not accounting for thermalization losses of other particles) can differ significantly between the three ejecta components. Consistent with previous studies \citep{Wanajo2014a,Kullmann2022a}, the heating rate in the dynamical ejecta follows closely the canonical rate of $10^{10}\times(\tpm/ 1\,\mathrm{d})^{-1.3}$\,erg\,\ginv\,\sinv, showing only weak bumpy features in the considered time interval of $0.1\,\mathrm{d}\la\tpm\la 100\,\mathrm{d}$. In the BH-torus ejecta, the heating rate exhibits slightly more pronounced features, with the bump at around $\tpm\sim 10\,$d being mainly produced by 2nd-peak elements ($^{132}$Te and $^{132}$I; cf.\,\citealp{Metzger2010c,Kullmann2022a}). In the NS-torus ejecta, where the r-process operates least efficiently, the heating rate is somewhat ($\sim 10\text{--}50\,\%$) lower than the heating rates in the other ejecta components at early times ($\tpm\la 2 \,$d) but afterwards increases relative to the others and eventually exceeds them by factors of several. This late-time increase is related to iron-group elements, namely (at times $1\,$d$\la\tpm\la 10\,$d) to $\beta^{-}$-decays of $^{72}$Ga and $^{66}$Ni, and electron capture of $^{56}$Ni, as well as (at times $\tpm\ga 10\,$d) to $\beta^+$-decay and electron capture of $^{56}$Co (cf. \citealp[][]{Wu2019c}).

We observe a good agreement between both nuclear networks, A and B (see panels~(d),~(g),~and~(i) of Fig.~\ref{fig:abund}), which suggests that uncertainties related to the network code do not affect the overall interpretation of our results concerning the relative role of each ejecta component. The main deviation between both networks is seen in the production of actinides, which remains sensitive to the difficult treatment of fission \citep[e.g.][]{Goriely2015a}.

\subsection{Comparison with AT2017gfo}\label{sec:comp-kilon-with}

\begin{figure*} 
  \includegraphics[width=0.99\textwidth]{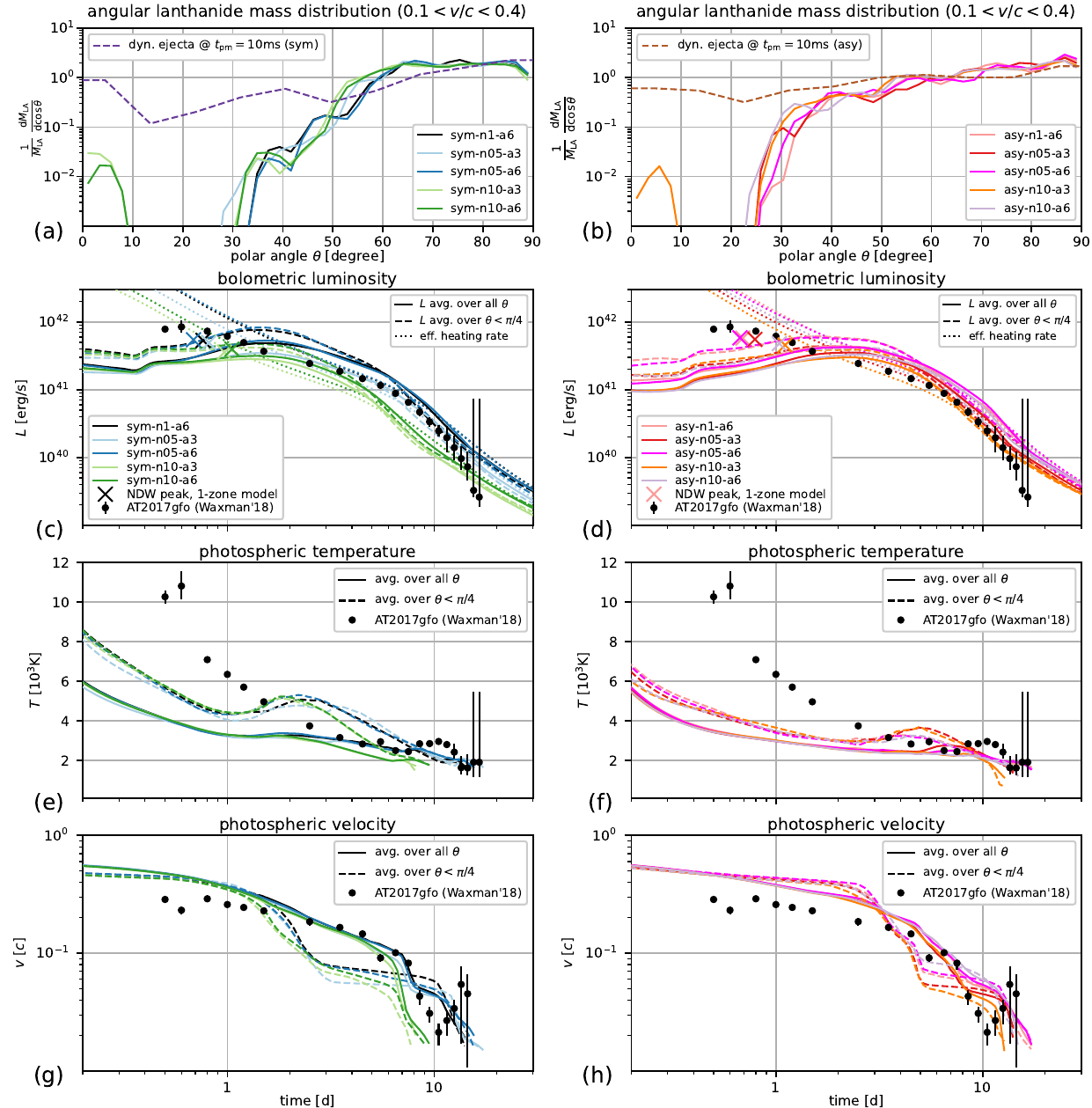}
\caption{Differential mass of lanthanides and actinides per solid angle along the polar angle (panels~(a),~(b)), bolometric, isotropic-equivalent luminosity and effective (i.e. including thermalization as in \citet{Rosswog2017b}) heating rate (panels~(c),~(d)), photospheric temperature (panels~(e),~(f)), and photospheric velocities (panels~(g),~(h)). Left (right) column shows plots for all models based on the symmetric (asymmetric) binary mass configuration. Plots in three bottom rows share the same x-axis and solid (dashed) lines denote quantities averaged over the entire sphere (over solid angles with $\theta<\pi/4$), while black circles denote data observed in AT2017gfo (from \citealp{Waxman2018a}). The crosses in panels~(c)~and~(d) denote peak-emission properties obtained from one-zone estimates \citep{Metzger2019a} using the mass, average velocity, and average opacity of all NS-torus ejecta with $Y_e>0.3$. Nucleosynthesis-related properties were obtained from network A.
  \label{fig:kilo}}
\end{figure*}

We first consider spherically averaged KN observables and discuss the viewing-angle dependence afterwards. For most models, the bolometric light curve (Fig.~\ref{fig:kilo}, panels~(c),~(d)) reaches peak emission after about $1\text{--}3$\,d, with luminosities of few$\,\times 10^{41}$\,erg\,\sinv well in the ballpark of AT2017gfo, and thereafter monotonically declines. Similarly to what was observed in AT2017gfo, we notice a shoulder-like feature around $\tpm\sim 5\text{--}8$\,d, which in our models is connected to the diffusion wave \citep{Waxman2018a}, i.e. the sudden release of accumulated radiation energy at the time when most of the ejecta become optically thin. At early times, $\tpm\la 1$\,d, our luminosities are lower than AT2017gfo by factors of $3\text{--}6$. Analogously, the photospheric temperatures and velocities\footnote{Both quantities are respectively computed as in Eqs.~(28) and~(29) of \cite{Just2022a}.} agree relatively well with the observation at times $\tpm\ga 2\text{--}5\,$d, but are slightly too cold and too fast, respectively, at earlier epochs.

Given the substantial anisotropy of the ejecta (e.g. panels~(a),~(b) of Fig.~\ref{fig:kilo}), we expect that our models show a strong viewing-angle dependence of the KN, e.g. in contrast to models based only on the dynamical ejecta \citep[e.g.][]{Just2022a, Collins2023a}. Since AT2017gfo was viewed from a polar angle, $\theta$, close to the rotation axis \citep[e.g.][]{Mooley2022a}, we plot the KN observables averaged over $\theta<\pi/4$ only (dashed lines in panels~(c)-(h)). The polar emission is characterized by about a factor of $2$ higher isotropic-equivalent luminosities and $20\text{--}40\,\%$ higher temperatures at $\tpm\ga 1\,$d, suggesting that a relatively larger contribution of emission is now stemming from the polar NDW, which has a lower opacity than the other ejecta components (cf. panel~(f) of Fig.~\ref{fig:cont}). At $\tpm \la 1\,$d, the polar light curves do not, similarly as the spherically averaged ones, reach the high fluxes observed in AT2017gfo.

One may wonder why the early peak of AT2017gfo is poorly reproduced by our models, despite the fact that the NDW properties (with mass $\sim$ 0.01-0.02\,$\Msun$, $Y_e\sim 0.4$, and velocity $\sim$ 0.2\,$c$ for models with $\nvis<10$) fulfill, at least marginally and better so for higher NS viscosity, the required conditions derived from fits to AT2017gfo based on spherical Arnett-type models \citep{Villar2017a, Smartt2017s}. We suspect one reason to be the circumstance that in contrast to an Arnett model, our NDW is not expanding spherically but rather conically, being partially shielded by lanthanide-rich dynamical ejecta both sideways and, at $v\ga 0.6\,c$, radially. The NDW photosphere visible to an observer near the pole is therefore not a sphere but subtends the smaller solid angle of a cone, and the photon fluxes released from the NDW are reduced by lanthanide-curtaining \citep{Kasen2015}. 

The ejecta properties in our current models may also be in tension with spectroscopic models of AT2017gfo \citep{Domoto2021a,Gillanders2022a,Vieira2023a} that, at least for early epochs ($\tpm\la 2\,$d), predict the line-shaping region surrounding the photosphere to be nearly free of lanthanides and heavier elements. This condition seems to be violated by the current models, in which for velocities $v/c\ga 0.1$ the lanthanide-free polar NDW component is embedded in lanthanide-rich dynamical ejecta (cf. green contours in panel~(e) of Fig.~\ref{fig:cont} showing radial photospheres that estimate the location of origin for radiation emitted at given times). Moreover, the very recent finding by \citet{Sneppen2023b} that the photosphere and strontium distribution may have been spherically symmetric to very high degree is difficult to reconcile with the anisotropic outflow structure seen in the current models.

Despite the inability to reproduce specific observational features, the fact that our models produce light curves with roughly the right brightness and decay timescales and overall similar features as the observations is certainly very promising and supports the possibility that a delayed-collapse merger akin to those investigated in this study was observed in AT2017gfo.

\section{Discussion}\label{sec:discussion}

A major unknown in merger models containing meta-stable NS remnants is the effective viscosity produced by MHD effects inside the NS. Compared to purely hydrodynamic systems, viscous merger remnants are able to tap the large reservoir of rotational and gravitational energy in the system and partially convert it to thermal energy. Since neutrino emission rates grow with high powers of the temperature, viscous merger remnants are therefore stronger sources of neutrinos than non-viscous remnants, implying that their NDWs are expected to be more powerful compared to non-viscous rotating NSs or to non-rotating proto-NSs (of which the NDWs have been extensively studied in the past; e.g. \citealp{Hudepohl2010a,Fischer2010}). Using for the first time an energy-dependent neutrino-transport scheme, our simulations confirm this expectation\footnote{A strong impact of viscosity on the NDW was also reported by \citet{Fujibayashi2017a} and \citet{Fujibayashi2018a}, who adopted an energy-independent neutrino treatment.} and demonstrate that NDWs in viscous NS remnants can be as massive as a few percent of $\Msun$ and exhibit velocity distributions reaching up to $v/c\sim 0.5\,c$. Judging from this result, mechanisms invoking genuinely magnetically-driven outflows \citep{Metzger2018b,Mosta2020c,Ciolfi2020s,Shibata2021c} may not be necessary in order to explain the early, blue component of AT2017gfo.

The adopted viscosity scheme is a parametrization of actually more complex physics. Our models aim at bracketing this uncertainty by generalizing the original Shakura-Sunyaev viscosity and introducing the viscosity parameter $\nvis$ that effectively regulates the viscosity just inside the NS. Despite a significant impact on the BH-formation time, $\tcol$, of about one order of magnitude, the total ejecta masses only vary by a factor of $\sim 2$ (in asymmetric merger models even less). Moreover, we find that the nucleosynthesis pattern is relatively robust with variations of $\nvis$, and the KN light curve varies only moderately. Although the absolute values of the total ejecta mass depend on the viscosity, we find for our (admittedly small) set of models that the total ejecta mass is systematically increased for asymmetric binaries, independently of the chosen set of viscosity parameters. This implies that for two observations with similar chirp mass the mass ratios can be related to each other, possibly allowing mass-ratio constraints based on the KN properties.

If we assume that our models are representative, the systematic underproduction found for $A>140$ elements would imply that mergers with intermediate (and probably also long, as suggested by \citealp{Fujibayashi2020b}) NS-remnant lifetimes could not be main r-process sites, but would likely be dominated by events not overproducing light relative to heavy elements, such as prompt or shortly-delayed collapse scenarios \citep{Just2015b,Fujibayashi2023a}.

As for the kilonova, our results suggest the considered delayed-collapse scenario to be a viable progenitor for GW17087/AT2017gfo, based on the overall good agreement of the trends seen in the bolometric light curve, photospheric temperature, and photospheric velocity at times $\tpm\ga 1\text{--}3\,$d. We find three features in which our models agree less well with observational analyses of the early ($\tpm\la$ few days) data from AT2017gfo, namely too faint emission, a non-spherical photosphere, and partial enrichment of the photospheric region by lanthanides (cf. Sect.~\ref{sec:comp-kilon-with}). We stress, however, that since radiative transfer calculations using detailed atomic-data based opacities are not yet available for delayed-collapse hydro models, our approximate KN modeling remains a non-negligible source of uncertainty. 

Since our set of models is not exhaustive, other combinations of binary masses, nuclear EOS, and viscosity parameters may yield better agreement with AT2017gfo. For instance, all three of the aforementioned inconsistencies could possibly be mitigated in cases where only a small amount of dynamical ejecta is launched such that the hot NDW is able to expand in a nearly spherical manner above and below the equatorial plane. Alternatively, it may also be possible that the anisotropic outflow structure seen in our present models is generic for delayed-collapse scenarios with significant NDW components. In this case one would expect that systems with shorter NS-remnant lifetimes, i.e. smaller binary masses, would yield generally more spherical ejecta distributions (cf. panels~(a)~and~(b) of Fig.~\ref{fig:kilo}) dominated by dynamical ejecta (for velocities $v/c\ga 0.1$). The two competing scenarios for the relative role of NDWs (i.e. increase vs decrease of sphericity with NS-remnant lifetime) could be distinguished by future kilonova obervations from systems with different binary masses relative to AT2017gfo using the P-Cygni method developed by \citet{Sneppen2023b}. At any rate, since the degree of sphericity in either of the two aforementioned scenarios depends on the lifetime, the P-Cygni method -- in combination with end-to-end hydrodynamical models that capture all ejecta components -- could be a powerful, new tool for constraining binary properties and the nuclear EOS from kilonova observations.

We finally point out that our models, despite featuring a coherent end-to-end modeling strategy, still carry non-negligible uncertainties connected to, e.g., the approximate treatment of GR, turbulent viscosity, and neutrino transport, the omission of magnetic fields, and the simplified KN physics. Moreover, our models are missing some physics ingredients that are potentially relevant to the KN problem, such as jets \citep[e.g.][]{Nativi2021d}, neutrino flavor conversion \citep[e.g.][]{Just2022k}, or non-axisymmetric NS oscillation modes that could produce additional ejecta components \citep{Nedora2019y}.

\acknowledgements We are grateful for inspiring discussions with Ninoy Rahman and Brian Metzger, and to the anonymous referee for comments that improved the manuscript. OJ, VV, and AB acknowledge support by the European Research Council (ERC) under the European Union's Horizon 2020 research and innovation programme under grant agreement nr 759253. VV and AB acknowledge support by Deutsche Forschungsgemeinschaft (DFG, German Research Foundation) - Project-ID 138713538 - Sonderforschungsbereich (Collaborative Research Center) SFB 881 (“The Milky Way System”, subproject A10). ZX and GMP acknowledge support by the ERC under the European Union’s Horizon 2020 research and innovation program (ERC Advanced Grant KILONOVA nr 885281). OJ, ZX, AB and GMP acknowledge support by the DFG - Project-ID 279384907 - SFB 1245. OJ, ZX, AB, GMP, and TS acknowledge support by the State of Hesse within the Cluster Project ELEMENTS. SG is F.R.S.-FRNS research associate. This work has been supported by the Fonds de la Recherche Scientifique (FNRS, Belgium) and the Research Foundation Flanders (FWO, Belgium) under the EOS Project nr O022818F and O000422. JG acknowledges support from the European Research Council (MagBURST, grant agreement nr 715368). HTJ is grateful for support by the DFG through SFB-1258$\,$--$\,$283604770 ``Neutrinos and Dark Matter in Astro- and Particle Physics (NDM)'' and under Germany's Excellence Strategy through Cluster of Excellence ORIGINS (EXC-2094)-390783311. OJ and ZX acknowledge computational support by the VIRGO cluster at GSI, and OJ by the HOKUSAI computing facility at RIKEN. The nucleosynthesis calculations benefited from computational resources made available on the Tier-1 supercomputer of the F\'ed\'eration Wallonie-Bruxelles, infrastructure funded by the Walloon Region under the grant agreement nr 1117545.

\appendix

\section{Neutrino grid resolution in SPH simulations}\label{sec:resol-depend-sph}

\begin{figure}
    \centering
    \includegraphics[trim=0  0  0  0,clip,width=0.95\textwidth]{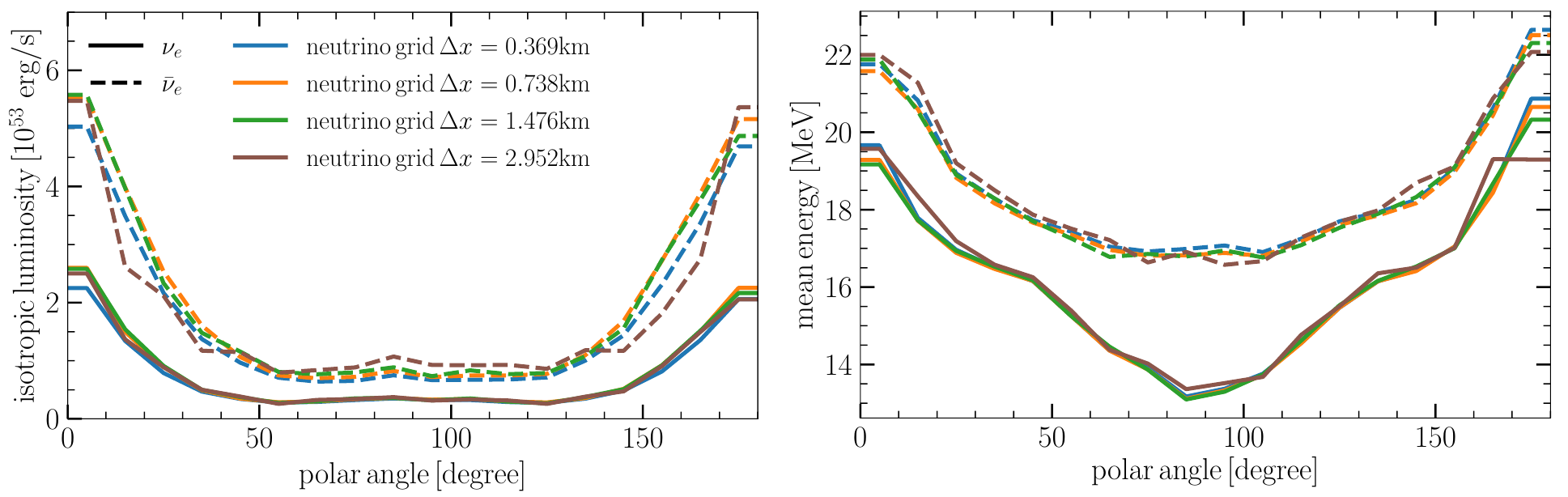}
    \caption{Isotropic-equivalent luminosities (left) and mean energies (right) as functions of polar angle for electron neutrinos (solid lines) and electron antineutrinos (dashed lines) resulting for different grid widths $\Delta x$ of the uniform, cartesian grid on which the ILEAS neutrino scheme is solved in the SPH merger simulations.}
    \label{fig:lumnu}
\end{figure}

The ILEAS scheme coupled to the SPH solver, which is used to simulate the merger phase until mapping to the ALCAR code at $\tmap=10\,$ms, computes the source terms related to neutrino emission and absorption on a uniform cartesian grid. In this appendix we briefly test the dependence of basic neutrino quantities on the grid resolution, to which end we pick a snapshot of our symmetric merger model at $\tpm\approx 5\,$ms and run only the ILEAS scheme on it keeping all hydrodynamic quantities fixed. Figure~\ref{fig:lumnu} depicts the resulting polar-angle dependent isotropic-equivalent luminosities, $L=4\pi r^2 F_r$ (with radial neutrino flux $F_r$), and mean energy, $\langle \epsilon\rangle$, measured at $r=100\,$km for various cases of the grid resolution $\Delta x$. The resolution dependence turns out to be mild, suggesting that the grid width of $\Delta x=0.738\,$km adopted in our dynamical models is small enough to ensure grid-related discretization errors to be subdominant (at least considering the early post-merger phase until $\tpm=10\,$ms when the surface of the NS remnant is still relatively hot and the density gradient rather shallow).

\section{Mass-ratio dependence of angular momentum in the NS remnant}\label{sec:mass-ratio-depend}

\begin{figure}
  \centering
  \includegraphics[width=0.99\textwidth]{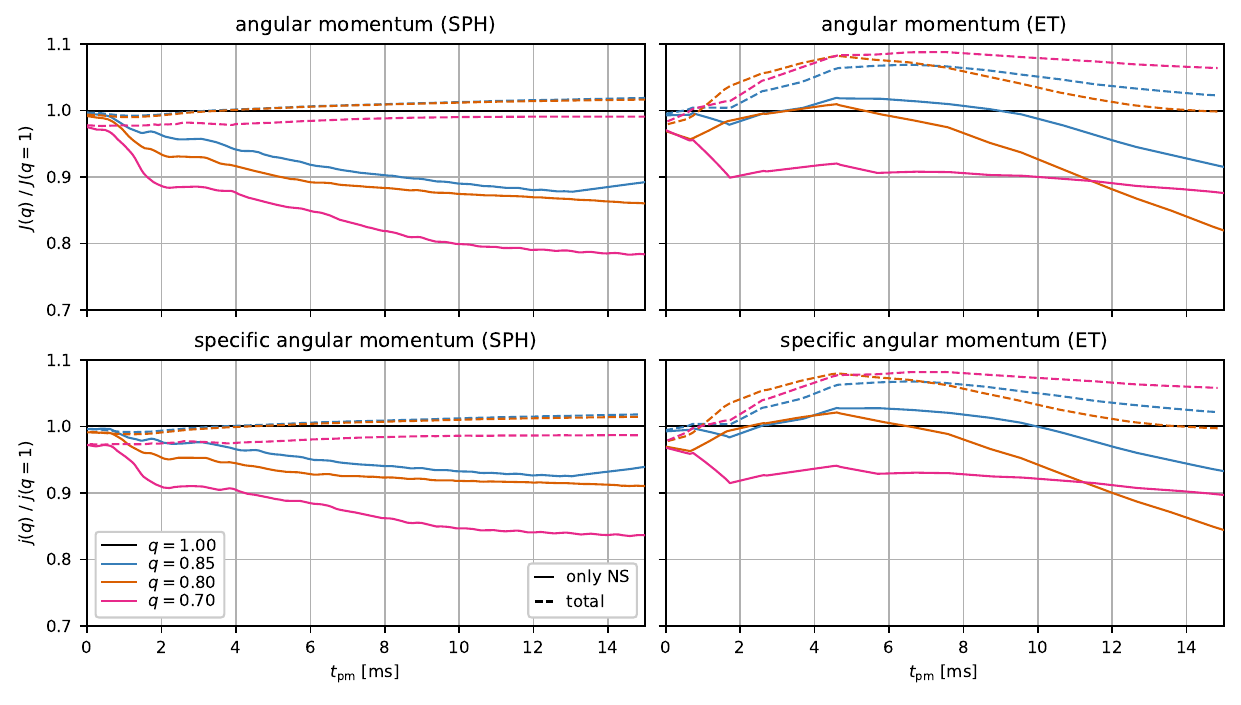}
  \caption{Angular momentum (top) and specific angular momentum (bottom) as functions of post-merger time for different binary mass ratios $q$, normalized by the corresponding values of the $q=1$ case, for simulations performed with our SPH code (left) and the Einstein Toolkit (ET; right). Solid lines take into account only NS material (i.e. regions where the density $\rho>10^{12}\,$g\,cm$^{-3}$) and dashed lines account for all matter in the integrals of Eqs.~(\ref{EqJz}). Note that since $\dd J/\dd t<0$ at all plotted times, values of $J(q)/J(q=1)$ greater (smaller) than unity indicate a slower (faster) decline of $J$ for given $q<1$ compared to the $q=1$ case.}
  \label{fig:Jzremnant}
\end{figure}

As pointed out in Sect.~\ref{sec:life-time-dependence}, we suspect the shorter NS lifetimes seen for our asymmetric models to be a consequence of the smaller amount of angular momentum carried by the high-density NS compared to the symmetric case. Here we show that this tendency -- i.e. smaller angular momentum carried by NS remnants of more asymmetric mergers -- is not only fulfilled for the two SPH simulations in the main part of this study, but also supported by another set of SPH simulations adopting a different total mass ($M_1+M_2=3\,\Msun$), different mass ratios, $q=M_1/M_2$, and different nuclear EOS (MPA1, \citealp{Muther1987a}), as well as by a similar set of models simulated with an entirely independent GR-hydro code, namely the Einstein Toolkit (ET, \citealp{Etienne2021a}). In contrast to the SPH simulations, the ET does not assume the conformal-flatness condition to approximate GR but solves the full GR equations. The numerical setup is the same as that described in \cite{Soultanis2022a}.

In Fig.~\ref{fig:Jzremnant} we compare for different values of $q$ the angular momentum, $J$, as well as the specific angular momentum, $j=J/M_b$ (with baryonic mass $M_b$), each normalized to the corresponding instantaneous value of the $q=1$ configuration. The angular momenta and masses are computed as (ignoring for this qualitative analysis non-compact space-time contributions to $J$):
\begin{subequations}\label{EqJz}
\begin{align}
  J_{\mathrm{NS/tot}}  &= \int \dd^3x~\rho_* \left(x\hat{u}_y-y\hat{u}_x\right) \, , \\
  M_{b,\mathrm{NS/tot}}&= \int \dd^3x~\rho_*  \, ,
\end{align}
\end{subequations}
where $\rho_*=\sqrt{\gamma}\rho W$ with the determinant of the spatial metric $\gamma$, Lorentz factor $W$, and $\hat{u}_i=h u_i$ with the specific enthalpy $h$ and the spatial component of the covariant fluid four-velocity $u_i$. As can be seen from the solid lines in Fig.~\ref{fig:Jzremnant}, if only the angular momentum carried by the NS remnant, $J_{\mathrm{NS}}$, is considered (by restricting the integration in Eq.~(\ref{EqJz}) to regions with $\rho>10^{12}\,$g\,cm$^{-3}$), both sets of simulations support the tendency of a faster reduction of $J_{\mathrm{NS}}$ and $J_{\mathrm{NS}}/M_{b,\mathrm{NS}}$ for lower $q$, although in the ET models in a somewhat less pronounced manner than in the SPH models. On the other hand, if one considers the total angular momentum, $J_{\mathrm{tot}}$ (obtained from integration over the entire computational domain in Eq.~(\ref{EqJz}); cf. dashed lines in Fig.~\ref{fig:Jzremnant}), one obtains rather the opposite tendency, namely a more slowly declining $J_{\mathrm{tot}}$ for lower $q$. In these purely hydrodynamic models the total angular momentum can change only due to the emission of gravitational waves (assuming that numerical non-conservation errors are small). Thus, the $J_{\mathrm{tot}}$ evolution likely reflects the fact that for a given total mass the post-merger gravitational-wave emission in equal-mass systems is stronger and therefore gravitational waves more efficiently reduce the $J_{\mathrm{tot}}$ than in asymmetric mergers \citep[e.g.][]{Kiuchi2020a}. Note, however, that very asymmetric configurations initially contain slightly smaller $J_{\mathrm{tot}}$ at the time of merging.

Thus, the $q$-dependence of the angular momentum remaining in the NS remnant is mainly shaped by two counteracting effects, namely gravitational-wave emission (which is less efficient for low $q$ values than for $q=1$) and redistribution of angular momentum into the low-density torus (which is more efficient for low $q$). Our results suggest the second effect to be stronger than the first one, leading ultimately to lower values of $J_{\mathrm{NS}}$ for $q<1$ at $\tpm\sim 10-15\,$ms. We stress, however, that our analysis is rather tentative at this point and that the above characteristics, as well as their consequences for the lifetime of the NS remnant, have to be tested and explored more systematically using models evolved for a longer time and with more realistic physics ingredients (i.e. neutrinos, magnetic fields) as well as including a detailed analysis of the impact of numerical discretization errors. In fact, we suspect that mainly numerical effects (and to a lesser extent the different treatment of GR) are the main reason for the differences between the SPH models and the ET models in the current test. The relatively high numerical viscosity of the SPH models may dampen post-merger oscillations, and therefore reduce angular-momentum losses by gravitational-wave emission, more strongly than in the ET models, which would explain the more pronounced reduction of $J_{\mathrm{NS}}$ with lower $q$ in the SPH models. This is suggested by simulations performed with our recently developed moving-mesh code employing CFC and the same gravitational-wave back-reaction scheme \citep{Lioutas2022o}, where we find the angular-momentum loss to be quantiatively more comparable to full-GR static-mesh simulations, corroborating that CFC is not the main reason for the quantitative differences between the SPH and ET results in Fig.~\ref{fig:Jzremnant}.

\section{Rotation profile in the NS remnant}\label{sec:rotat-prof-remn}

\begin{figure}
  \includegraphics[width=\textwidth]{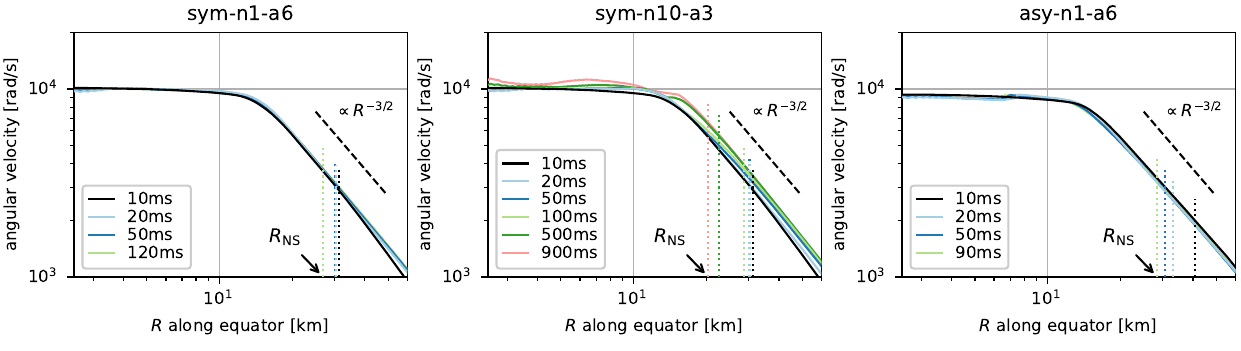}
  \caption{Angular velocity $\Omega = v_\phi/R$ as a function of cylindrical radius $R=r\sin\theta$ along the equator at the post-merger times, $\tpm$, indicated in the legends for the three models sym-n1-a6, sym-n10-a3, and asy-n1-a6. Dotted vertical lines indicate for each time the location of the NS surface, i.e. the radius where $\rho=10^{12}\,$g\,cm$^{-3}$. Dashed lines indicate the slope of profiles proportional to $R^{-3/2}$.}
  \label{fig:angvel}
\end{figure}

The radial profiles of the angular velocity, measured at the time of mapping from the 3D merger models to the 2D post-merger models, $\tpm=\tmap=10\,$ms, as well as for various later times, are shown in Fig.~\ref{fig:angvel} for three models. The inner core of the merger remnant is rotating nearly uniformly already at the time of mapping, while the profile in the surrounding disk corresponds to Keplerian rotation ($\propto r^{-3/2}$). These characteristics remain essentially unchanged throughout the entire evolution of the NS remnant. We note that other results in the literature exist that report somewhat different behavior shortly after the merger, namely a combination of a slowly rotating inner and a fast rotating outer core  \cite[e.g.][]{Hanauske2017a}, or a double-core structure surviving for a significantly longer time \citep{Lioutas2022o}. We suspect that these differences are connected to the different numerical discretization schemes adopted by the aforementioned models (SPH vs. cartesian grid vs. moving mesh, respectively), however, a detailed understanding of these differences has yet to be obtained.

\section{Mapping of dynamical ejecta}\label{sec:mapp-dynam-ejecta}

\begin{figure}
  \newcommand\heiappa{0.26}
  \newcommand\heiappb{0.26}
    \includegraphics[trim=10  10 7  0,clip,width=0.511\textwidth,height=\heiappa\textheight]{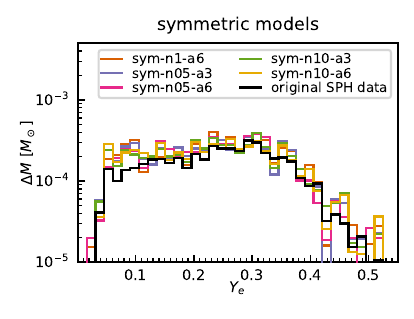}
    \hspace{0.02\textwidth}
    \includegraphics[trim=36  10 7  0,clip,width=0.437\textwidth,height=\heiappa\textheight]{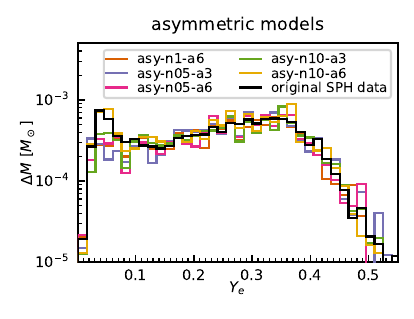} \\
    \includegraphics[trim=62  0 37 50,clip,width=0.523\textwidth,height=\heiappb\textheight]{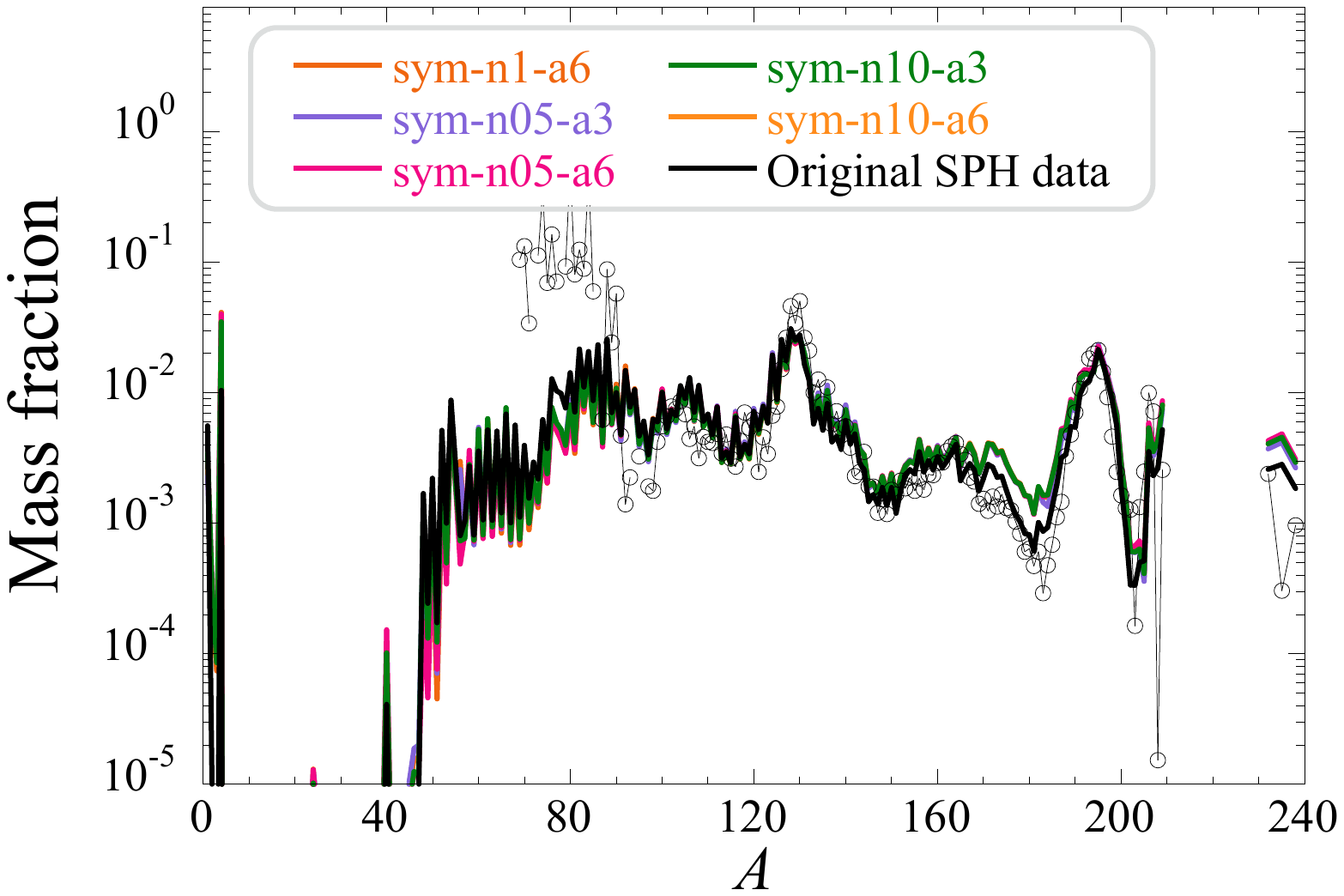}
    \hspace{0.008\textwidth}
    \includegraphics[trim=164 0 37 50,clip,width=0.451\textwidth,height=\heiappb\textheight]{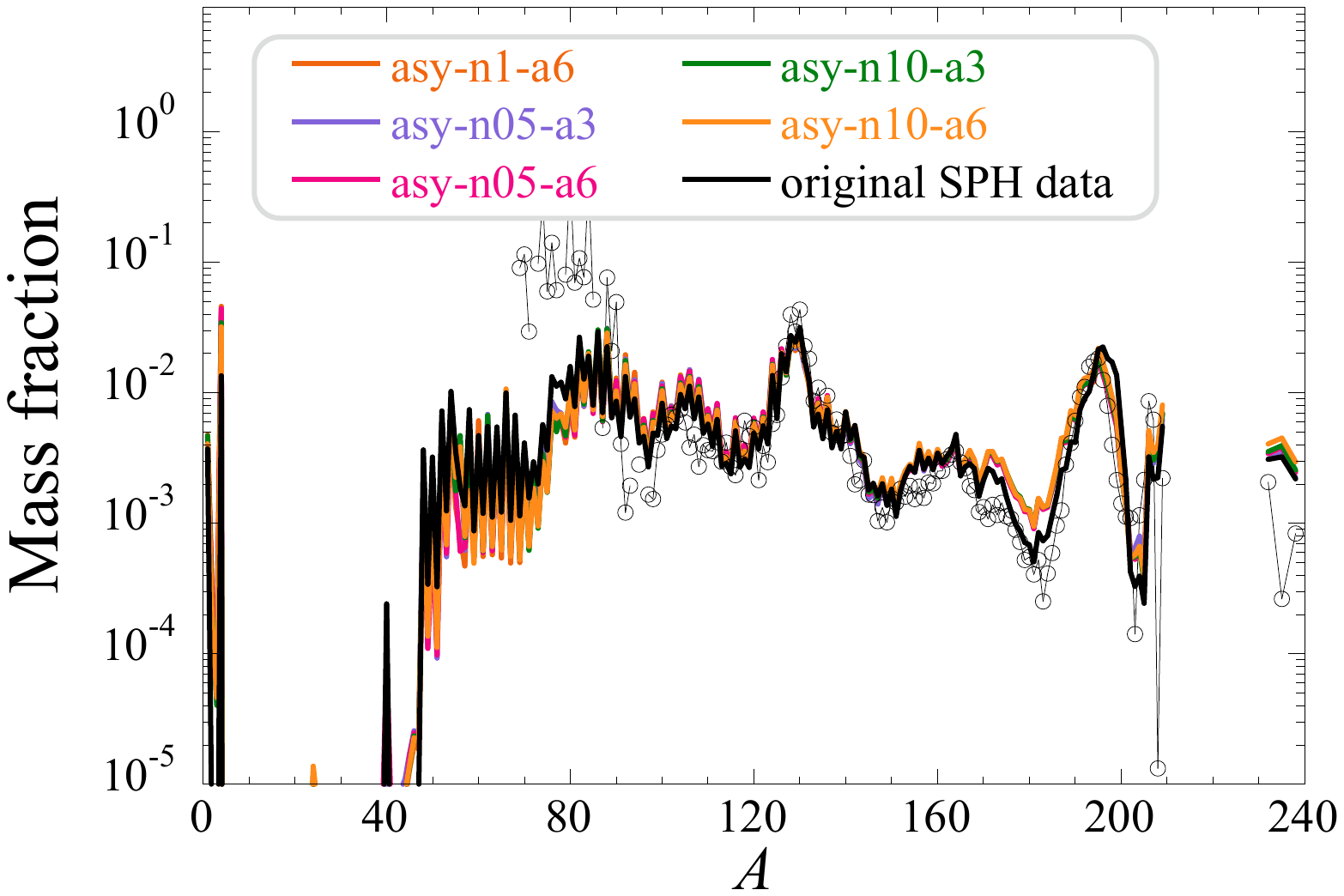}
    \caption{$Y_e$ histograms (top row) and nucleosynthesis yields (bottom row) of the dynamical ejecta as resulting in our end-to-end models (which combine data from 3D SPH and 2D grid simulations) compared to the corresponding properties resulting just for the SPH simulations (black lines) for all models based on the symmetric (left column) and asymmetric (right column) binary mass configurations.}
    \label{fig:dynmap}
\end{figure}

When constructing the outflow trajectories via backward time integration, special care must be taken to ensure that the distribution of thermodynamic properties (most importantly of $Y_e$) in the dynamical ejecta remains consistent with that of the original merger simulations. This is because variations of fluid properties on small spatial scales or along the azimuthal direction get averaged out by the mapping from the 3D SPH configuration to the 2D grid at $\tmap=10\,$ms. In order to approximately retain the $Y_e$ pattern of the SPH simulations we construct the dynamical-ejecta trajectories as follows: After backwards integration from $\tpm=100\,$s to $\tmap$, all trajectories fulfilling $r(\tmap)>250\,$km are split into five trajectories that differ only by their mass and $Y_e$. The mass- and $Y_e$-values for these five trajectories are taken from the five SPH particles of the corresponding merger model with the closest locations to the original post-merger trajectory at that time. The masses of these SPH particles are normalized such that their sum equals the mass of the post-merger trajectory. In the case that for these trajectories the temperature already dropped below $10\,$GK at $\tmap$, their expansion history at earlier times is taken directly and consistently from the SPH simulation. The resulting $Y_e$ histograms and abundance yields are compared with those of the original SPH data in Fig.~\ref{fig:dynmap}. We find overall good consistency between both data sets, but also noticeable differences for post-merger models based on the same SPH model, particularly at low $Y_e$ values (e.g. the $Y_e\approx 0.05$ peak in the asymmetric models). However, these differences are not necessarily connected only to sampling errors (i.e. errors related to the aforemented mapping at $\tmap$ as well as to the finite number of post-merger tracers), but could to some extent also be caused by different late-time ($\tpm>\tmap$) behavior: While the post-merger simulations capture the hydrodynamic evolution of the ejecta far beyond $\tmap$ -- including effects such as fallback or interaction with other ejecta components, which all can be sensitive to the viscosity -- the plotted SPH data assumes spherical, adiabatic expansion to extrapolate beyond $\tmap$. Nevertheless, despite the approximate mapping and the different assumptions at late times, the global pattern and most relevant features of the nucleosynthesis yields agree very well.

\section{Origin of polar outflow from NS remnant}\label{sec:relat-import-neutr}

\begin{figure}
  \includegraphics[trim=0  0 0 0,clip,width=0.99\textwidth]{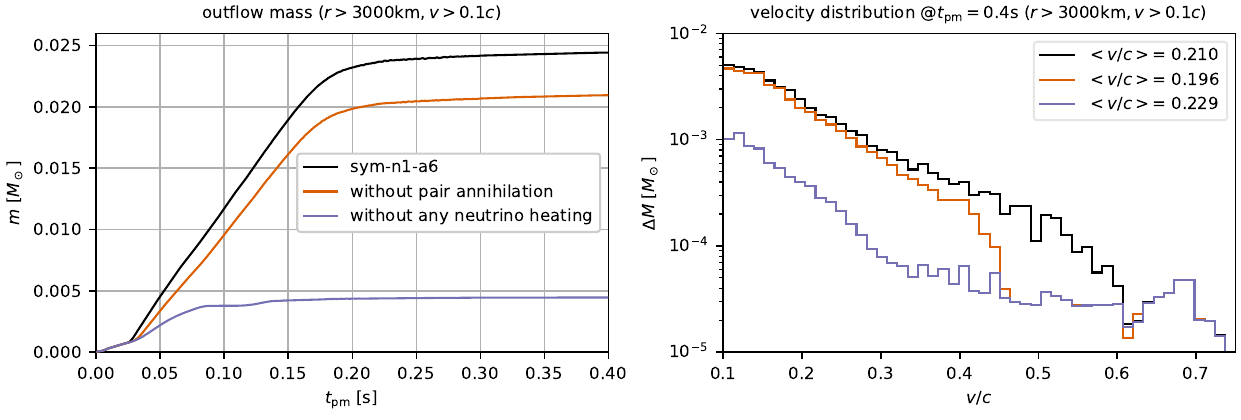}
  \caption{Left panel: Mass of material with radii $r>3000\,$km and velocities $v/c>0.1$ as function of time compared between model sym-n1-a6 and a corresponding model without neutrino pair annihilation as well as another model where net neutrino heating is neglected entirely. Note that $m(t)$ saturates significantly later than the time of BH formation ($\tcol\approx 122\,$ms) because of the time needed by the ejecta material to travel from the NS surface to $r=3000\,$km. Right panel: Mass-velocity distribution measured for the same models and ejecta at $\tpm=0.4\,$s.}
  \label{fig:anniout}
\end{figure}

In the main text we argued that the polar outflow observed before BH formation is driven by neutrino heating without, however, explicitly backing this statement. We also did not discuss the role of neutrino pair annihilation for driving this outflow. In order to briefly address these aspects, we ran two additional simulations similar to model sym-n1-a6, one in which only heating due to neutrino pair annihilation is ignored and another one in which also heating due to neutrino-nucleon absorption is neglected. As for the technical implementation of these modifications, at each integration step we first compute all source terms as usual, but then set to zero all source terms corresponding to the aforementioned neutrino interactions (in both the hydro- and moment-equations) in regions where the density is smaller than $10^{11}\,$g\,cm$^{-3}$ and neutrino interactions would otherwise heat up the fluid. The results are shown in Fig.~\ref{fig:anniout}, which depicts the mass of all material with velocities $v/c>0.1$ ejected beyond $r=3000\,$km as well as the mass-velocity distribution of the same material at $\tpm=0.4\,$s, late enough to capture all fast ejecta from the NS remnant that collapses at $\tcol= 122\,$ms. Without any neutrino heating (purple lines) the mass ejected within the first few hundred milliseconds is about five times smaller than in the unmodified model sym-n1-a6 (black lines) and corresponds to just about the mass of the dynamical ejecta with $v/c>0.1$, demonstrating that neutrino heating is indeed the main driver of the fast, polar outflow. The relative impact of pair annihilation can be assessed when comparing with the orange lines, which reveal that the ejecta exhibit a slightly less extended high-velocity tail, reaching only up to $v/c\approx 0.45$ instead of 0.6, when pair annihilation is not taken into account. However, the total ejecta mass is reduced only by about $4\times 10^{-3}\,\Msun$ (corresponding to $\approx 5\,\%$ of the total ejecta mass for this model), suggesting that pair annihilation has only a small impact on r-process- and kilonova-related features of the NS-torus ejecta in our models.



\end{document}